\documentclass[twocolumn,10pt,pre,aps,showpacs]{revtex4-1}

\usepackage[dvips]{graphicx}
\usepackage[usenames]{color}
\usepackage[normalem]{ulem}
\graphicspath{{./pics/}}

\def\beq{\begin{equation}}
\def\eeq{\end{equation}}
\def\bea{\begin{eqnarray}}
\def\eea{\end{eqnarray}}

\begin{document}

\title{Comparison of electron-ion energy transfer in dense plasmas \\
       obtained from numerical simulations and quantum kinetic theory}

\author{J.\ Vorberger}
\affiliation{Max Planck Institut f\"ur die Physik komplexer Systeme, 01187 Dresden, 
             Germany}
\author{D.O.\ Gericke}
\affiliation{Centre for Fusion, Space and Astrophysics,
             Department of Physics, University of Warwick,
             Coventry CV4 7AL, United Kingdom}

\date{\today}

\begin{abstract}
We evaluate various analytical models for the electron-ion energy transfer
and compare the results to data from molecular dynamics (MD) simulations.
The models tested includes energy transfer via strong binary collisions,
Landau-Spitzer rates with different choices for the cut-off parameters in
the Coulomb logarithm, rates based on Fermi's golden rule (FGR) and
theories taking coupled collective modes (CM) into account. In search of
a model easy to apply, we first analyze different approximations of the
FGR energy transfer rate. Then we investigate several numerical studies 
using MD simulations and try to uncover CM effects in the data obtained.
Most MD data published so far show no distinct CM effects and, thus, can
be interpreted within a FGR or binary collision approach. We show that
this finding is related to the parameter regime, in particular the initial
temperature difference, considered in these investigations.
\end{abstract}

\maketitle

\section{Introduction}
Energy exchange between particles of different species and the subsequent 
equilibration of their temperatures plays a crucial role in many modern
experiments creating high-energy-density matter. A prime example is inertial
confinement fusion where intense lasers are employed to heat and compress
a deuterium-tritium pellet to extreme conditions
\cite{lindl,Glenzer:2010,Xu:2011,NIF_rep}. Plasmas with well separated electron
and ion temperatures are created in the laser spots, the ablator, and the
hydrogen pellet itself. In particular, the creation and propagation of the
burn wave in the fuel depends on electron-ion temperature equilibration as the 
main portion of the hydrogen needs to remain cold during compression. Further
heating is achieved by $\alpha$-particles from the hot spot that, however, 
transfer their energy almost entirely to the electrons. Fusion conditions are
thus only reached after sufficient energy transfer from the electronic to the
ionic subsystem.

Temperature relaxation is also very important for most experiments dedicated to
investigate warm dense matter because a fast energy input into the sample is
required to create such a state in the laboratory
\cite{Ng, Riley,french,Glenzer:2007,Garcia:2008,Pelka:2010,White:2012}. As the
energy is mostly delivered to one species and into special states, samples are
often driven into states far from equilibrium. After a short initial
equilibration within each species, the system can be characterised by 
different electron and ion temperatures. Interestingly, ultra-fast probing
\cite{Kritcher:2008,Barbrel:2012} allows nowadays to investigate dense plasmas
on the time scale of temperature equilibration (a few pico-seconds). Even
shorter fs-pulses of intense light as delivered by free electron lasers can
further increase the time-resolution or may be used to probe the initial 
relaxation as well \cite{Faustlin:2010,Chapman:2011}. Knowledge of the
temperature relaxation time is also important when determining equilibrium
properties like the equation of state or the structure in equilibrium as it
sets a minimum time between creation and probing. 

As experimental data for the electron-ion equilibration time are sparse, one
must often rely on theoretical and simulation results for many applications.
The theoretical investigations of electron-ion temperature equilibration can be 
grouped in two main strains: molecular dynamics (MD) simulations 
\cite{Dimonte:2008,Jeon:2008,Glosli:2008,Murillo:2008,
      Benedict:2009,Daligault:2009,Benedict:2012} 
and quantum kinetics  
\cite{Dharma:98,Hazak:2001,Gericke:2002b,Daligault:2008,Dharma:2008,
      Gregori:2008,Vorberger:2009,Vorberger:2010,Baalrud:2012}. 
Both approaches have their strengths and shortcomings: MD simulations are
inherently non-perturbative and take into account all channels of energy
transfer between the electrons and ions. However, they are based on classical
physics for both the electrons and the ions which strongly restricts their
range of applicability. Moreover, the use of electron-ion pseudo-potentials
designed to mimic aspects of the quantum behaviour raises basic questions. 

Quantum statistical theory, on the other hand, is mostly based on perturbation
theory with respect to the electron-ion interaction strength. Thus, it offers
rates that take into account either strong collisions or energy transfer via
collective modes. However, statistical theory can easily account for quantum
effects like diffraction and exchange during the scattering process as well as
Fermi statistics in degenerate systems.

Here, we conduct an extensive comparison of available MD data with results from
quantum statistical theory. First we investigate the electron-ion energy
transfer for parameters where strong binary collisions are important. In this
connection, we analyze parametrizations of the energy transfer rates in terms
of a modified Coulomb logarithm that is chosen to match results of a full
binary collision theory (T-matrix approach) \cite{Gericke:2002b} or data
from MD simulations. Then we evaluate the quality of different approximations
often applied when computing of the energy transfer within the FGR approach
and, identify parameter regions where these approximations are
applicable. The limits of applicability are particularly important when
searching for CM effects by comparing to energy transfer rates based on FGR.
Finally, we compare energy transfer rates based on binary collision theory,
Fermi's golden rule and the CM approach to results from MD simulations. We
find only small hints of CM effects in few of the MD data. The further analysis
shows that this finding is related to the small temperature difference
considered in most simulations. Moreover, some of the published MD simulations
were performed for parameters outside the truly classical regime which may 
contribute to the inconsistencies found.

\section{Binary collision approach}
The electron-ion energy transfer via binary collisions is traditionally
described in form of the Landau-Spitzer (LS) approach
\cite{Landau:1937,Spitzer:1967} valid for weakly coupled, classical plasmas.
The LS formula for the rate of energy transfer between electrons and ions
is expressed in terms of the Coulomb logarithm $\ln\Lambda$
\beq
Z_{ei} = \frac{k_B \left(T_i - T_e \right)}
            {\left(\frac{k_BT_e}{m_e}+\frac{k_BT_i}{m_i}\right)^{3/2}} \,
         \frac{4\sqrt{2\pi}n_e n_i Z_i^2 e^4}{m_em_i} \,
         \ln\Lambda \,.
\eeq
Here, we have the temperature $T_a$ of species $a=\{e,i\}$, the densities $n_a$,
ion charge state $Z_i$, the masses $m_a$, the elementary charge $e$, and
Boltzmann's constant $k_B$. The Coulomb logarithm contains the information
about the strength of the classical two-particle scattering process in the
plasma medium and is given by the upper and lower cutoff parameters. The
original LS approach, considering straight electron trajectories, yields
\beq
\ln\Lambda^0 = \ln\frac{b_{max}}{b_{min}} \,.
\label{lnl1}
\eeq
Many variations for the cutoffs $b_{min}$ and $b_{max}$ have been suggested.
More advanced expressions take into account hyperbolic electron orbits in the
Coulomb field of the ion and changes to the scattering process due to quantum 
diffraction effects in close orbits
\beq
\ln\Lambda^H = \frac{1}{2} 
               \ln\!\left(1 + \frac{b^2_{max}}{b^2_{min}} \right)
             = \frac{1}{2} 
                \ln\!\left(1 + \frac{\lambda_D^2}
                                    {\varrho^2+\lambda_{dB}^2}\right) \,.
\label{lsh}
\eeq 
Here, $\lambda_D \!=\! \sqrt{k_B T_e/(4\pi e^2 n_e)}$ is the classical
screening (Debye) length, $\varrho \!=\! Z_ae^2/k_BT_e$ denotes the distance of
closest approach in the collision, and $\lambda_{dB}^2 \!=\! 1/m_e k_B T_e$ is the
deBroglie wavelength defining the range of quantum effects.

One of the shortcomings of the LS approach is that it considers collisions in
Coulomb fields and incorporates screening by a cutoff only. Electron-ion
collisions in the screened (Debye) potential were considered in the full binary
collision approach that is based on the quantum Boltzmann equation
\cite{Gericke:2002b}. As the collision cross sections were obtained from
solutions of the Schr\"odinger equation \cite{Gericke:1999}, quantum
diffraction is also fully accounted for. The results were fit to the form
(\ref{lsh}), using the upper cutoff as a free parameter 
\bea
\ln\Lambda^{bc} =
         \frac{1}{2} \ln\!\left(1 + \frac{b_{up}^2}
                                         {\rho^2+\lambda_{dB}^2}\right)
\nonumber
\eea
with
\beq
b_{up} = \lambda_D \cdot
          \exp\left\{\frac{1.65 - 0.4 \cdot \ln\Lambda^H}
                          {(\ln\Lambda^H)^{0.64} + 1} \right\} \,.
\label{lst}
\eeq

\begin{figure}[t]
\includegraphics[width=0.48\textwidth,clip=true]{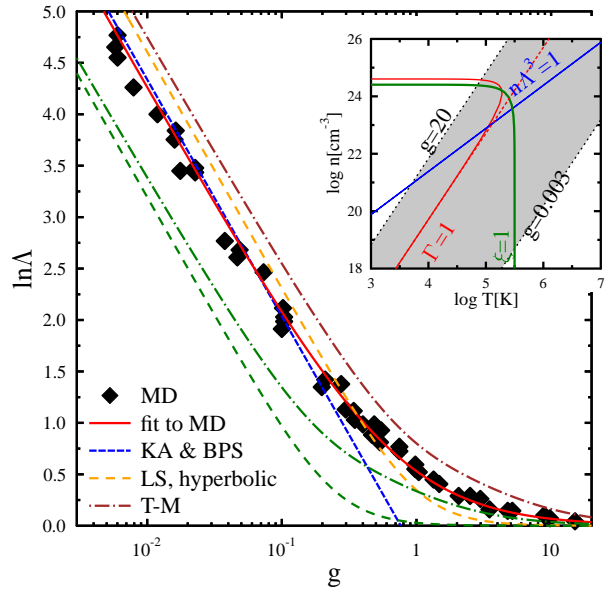}
\caption{(color online) Coulomb logarithm for classical systems as derived from
         MD simulations by Dimonte \& Daligault \cite{Dimonte:2008}
         and as predicted by various theories: KA \cite{KA:1963}, BPS \cite{BPS:2006}. 
	 The results are plotted
         versus the coupling parameter 
         $g \!=\! \varrho/\lambda_D \!=\! Ze^2/(k_B T \lambda_D)$ and
         for a temperature of $T \!=\! 5 \!\times\! 10^6\,$K. 
         The result from full binary collision theory (brown dash-dotted)
         and the LS curve (orange dashed) omit the deBroglie wavelength
         in the minimum cutoff of the Coulomb logarithm as it should be
         for purely classical systems. The green curves (dash-dotted for TM, dashed for LS)
	 illustrate the
         effect of the deBroglie wavelength in $b_{min}$.
         Quantum effects can be estimated by the parameters
         $\xi \!=\! \varrho/\lambda_{dB}$ and $n\Lambda^3$ (see text).
         \label{fig4}}
\end{figure}

We compare the formulae (\ref{lsh}) and (\ref{lst}) with Coulomb logarithms
extracted from classical MD simulations \cite{Dimonte:2008} in Fig.~\ref{fig4}.
The best fit to the MD data has the same functional form as Eq.~(\ref{lsh}):
$\ln\Lambda \!\sim\! \ln(1 \!+\! 0.7/g)$. The MD simulations were performed
for particles with the same charge which allowed for the application of pure
Coulomb interactions (no electron-ion pseudo-potential) in classical dynamics. 
To be consistent with a classical system like this, we omit the deBroglie
wavelength in the minimum cutoff of the Coulomb logarithm. The results of 
both Eqs.~(\ref{lsh}) \& (\ref{lst}) are in acceptable agreement with
the MD data for small coupling and even reproduce the correct behavior for high values of the 
coupling constant $g$ (strong interactions) reasonably well. Therefore, these 
MD data clearly show the effects of strong
binary collisions. However, CM effects (intentionally not included in the full
binary collision approach above) are not visible.

The effect of quantum diffraction can be estimated by including the deBroglie
wavelength in the minimum cutoff. We illustrate this effect with the green
curves in Fig.~\ref{fig4}. Clearly, the consideration of the deBroglie
wavelength considerably reduces the Coulomb logarithm and thus the
electron-ion energy transfer rate. The importance of diffraction can be
estimated by the Born parameter $\xi \!=\! \varrho/\lambda_{dB}$ that compares
the deBroglie wavelength with the Landau length, i.e. the mean of the closest 
approach possible between two particles at a certain temperature. For high 
temperatures, this parameter is small and diffraction
is thus important. For high densities/low temperatures, quantum degeneracy
limits the classical description. The parameter 
$n_e\Lambda_e^3 \!=\! n_e (2\pi\hbar^2 / m_e k_B T_e)^{3/2}$ relates the thermal
wavelength of the electrons to their mean distance. When this parameter
approaches unity, the electrons become degenerate prohibiting a classical
description. Contours of the Born and degeneracy parameters are shown in the
inset of Fig.~\ref{fig4} showing the applicability of classical MD simulations
in the density-temperature plane.

\section{Collective Excitations}
Collective excitations can strongly modify the electron-ion energy transfer
rates. To include this effect, one has to go beyond the binary collision 
approach. Considering longitudinal collective excitations in the electron and
ion subsystem and their mutual influence, one arrives at the CM expression for
the energy transfer rate \cite{Dharma:98,Vorberger:2010}
\beq
\label{coupled_mode_formula2}
Z_{ei}^{CM} = -4 {\cal V} \!\int\!\frac{d {\bf k}}{(2\pi)^3}
                 \!\int\limits_0^{\infty}\frac{d\omega}{2\pi\hbar} \,
                 \omega \, \Delta N
                 \frac{{\rm Im}\varepsilon_e({\bf k};\omega) \,
                       {\rm Im}\varepsilon_i({\bf k};\omega)}
                      {|\varepsilon({\bf k};\omega)|^2} \,.
\eeq
Here, $\Delta N_B \!=\! n_B^e(\omega,T_e) \!-\! n_B^i(\omega,T_i)$ denotes the
difference of Bose functions describing the occupation of electron and ion
modes. The coupling between the electron and ion modes is facilitated by the
dielectric function of the fully coupled electron-ion system
\bea
\label{DK}
\varepsilon({\bf q};\omega) =
          1 - \sum_a  V_{aa}(q) \, \Pi_{aa}^R({\bf q};\omega)
\eea 
expressed by the polarisation function of electrons/ions.

Independent collective excitations in the electron and ion subsystem are
described within the FGR approach. It can be obtained from the coupled mode
expression (\ref{coupled_mode_formula2}) by splitting the full dielectric
function in the denominator into a product of electron and ion contributions
\cite{Dharma:98,Vorberger:2010}
\bea
Z_{ei}^{FGR} &=& -4{\cal V} \!\int\!\frac{d{\bf k}}{(2\pi)^3}
                   \!\int\limits_0^{\infty}\frac{d\omega}{2\pi\hbar} \,
                   \omega \, \Delta N
\nonumber\\
             & & \times\mbox{Im}\varepsilon_{e}^{-1}(k;\omega) \,
                       \mbox{Im}\varepsilon_{i}^{-1}(k;\omega) \,.
\label{fgrfull}
\eea
Here, Im$\varepsilon_{aa}^{-1}$ are the spectral functions of the collective
excitation for the species $a$. The collective excitations are of Bosonic
character and, thus, are populated with a Bose distribution $n_B^a$ for
different temperatures $T_a$.

The FGR expression (\ref{fgrfull}) can be evaluated numerically. However, one
would like to avoid the double integration for many applications. This can be
achieved using the low $\omega$ expansions of the Bose function and the
dielectric response of the electrons 
\bea
n_B^a(\omega) &\approx& \frac{k_BT_a}{\hbar\omega}                     \\
\frac{\mbox{Im}\varepsilon_e^{-1}}{\omega} &\approx&
\frac{\mbox{Im}\varepsilon_e^{-1}}{\omega}\Big|_{\omega=0} = C(k)
\eea
which allows to perform the $\omega$-integral using the f-sum rule for the ions
\cite{Hazak:2001}.
The approximations above are often applicable due to the separation of the
electron and ion excitations in energy space. Moreover, the ion response
function effectively limits the $\omega$-integral in the energy transfer rate
to small frequencies. 

For the electron-ion energy transfer rate, we arrive at
\beq
Z_{ei}^{(1)} = k_B (T_i-T_e) \frac{4\pi Z_i^2e^2 n_i}{m_i}
               \int\!\frac{d{\bf k}}{(2\pi)^3} \,
               \frac{\mbox{Im}\varepsilon^{-1}_{ee}(k,\omega)}{\omega}\Big|_{\omega=0} \,.
\eeq
For arbitrary degeneracy, one can express the electron dielectric function in
the low frequency limit by static screening and the Fermi distribution for the
electrons. Then one obtains
\beq
Z_{ei}^{(2)} = k_B (T_i-T_e) \frac{4 Z_i e^4 n_e m_e^2}{\pi m_i}
               \int\limits_0^{\infty}\!dk \,
               \frac{k^3}{(k^2 + \kappa_e^2)^2} \; 
               f_e\!\!\left(\frac{k}{2},\mu\right)\!.
\label{fgr_ad}
\eeq
The screening parameter $\kappa_e$ is calculated via
\beq
\kappa_e^2 = \frac{4\pi e_e^2}{m_ek_B T_e \Lambda_e^3}
             \mbox{I}_{1/2}(\beta\mu_e) \,,
\eeq
where $\Lambda_e \!=\! \sqrt{2\pi \hbar^2/m_e k_B T_e}$ is the electron's
thermal wavelength and $I_{1/2}$ is the Fermi integral of order $1/2$.

In the limit of low electron degeneracy, the screening length becomes the
Debye length and the Fermi function a Boltzmann distribution. Then the energy
transfer rate simplifies to
\bea
Z_{ei}^{(3)} &=& k_B (T_i-T_e) \frac{16\sqrt{\pi m_e}n_e^2Z_ie^4}{m_i(2k_BT_e)^{3/2}}
\nonumber\\
             & & \times\left\{e^{a}\mbox{Ei}(a) - 1 
                              + e^{a} a\mbox{Ei}(a)\right\},
\label{fgr_ld}
\eea
with $a=\lambda_{dB}^2\kappa_e^2/8$ and
\beq
\mbox{Ei}(x)=\int\limits_x^{\infty}ds\,\frac{e^{-s}}{s}\,.
\eeq

For highly degenerate systems, we get from a Sommerfeld expansion of the
electron distribution function
\bea
Z_{ei}^{(4)}&=&k_B(T_i-T_e)\frac{4 n_e Z_i e^4 m_e^2}{\pi m_i}
\label{fgr_hd}\\
&&\times\Bigg\{\frac{1}{2}\left[\ln(4\mu_e+\kappa_e^2)
+\frac{\kappa_e^2}{4\mu_e+\kappa_e^2}
-\ln(\kappa_e^2)-1
\right]
\nonumber\\
&&\;\;\;\;\;\;+\frac{\pi^2k_B^2T_e^2}{6}
\left[\frac{12\mu_e}{(4\mu_e+\kappa_e^2)^2}
-\frac{64\mu_e^2}{(4\mu_e+\kappa_e^2)^3}
\right]\Bigg\}\,.\nonumber
\eea
In this case, the chemical potential $\mu$ is the Fermi energy of the electrons.

\begin{figure}[t]
\includegraphics[width=0.48\textwidth,clip=true]{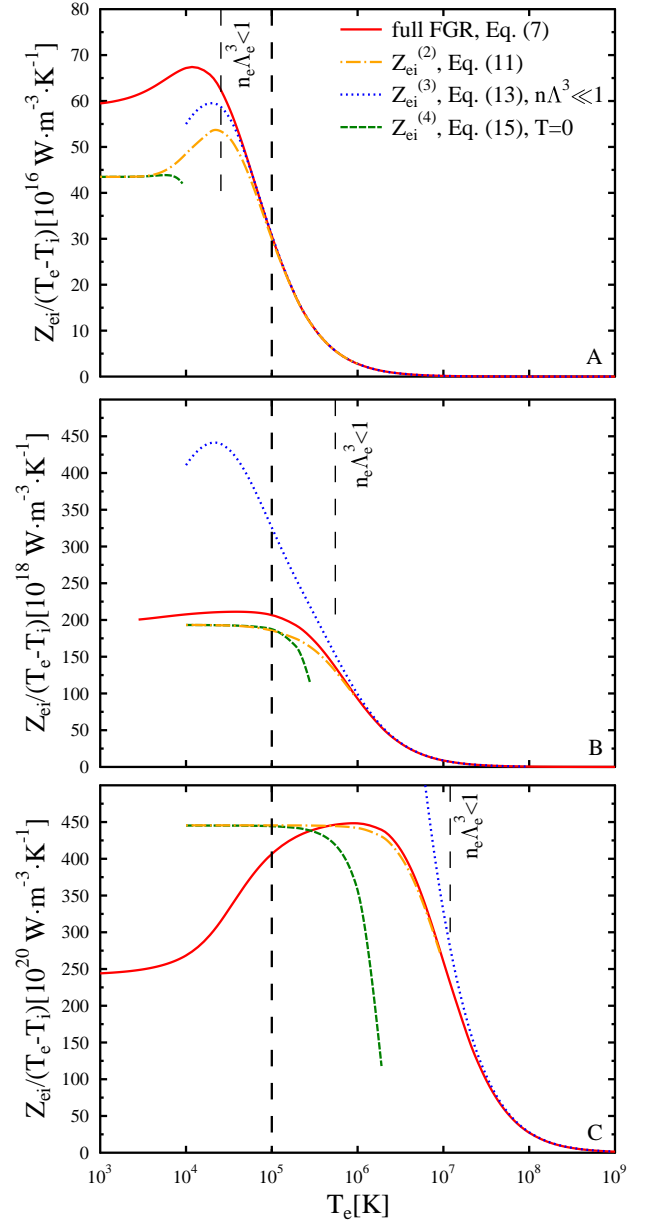}
\caption{(color online) Electron-ion energy transfer rates for a fully ionised
         hydrogen plasma in different approximation levels within
         the FGR description versus electron temperature.
         The plasma densities are:
         $n_e \!=\! 10^{22}$cm$^{-3}$ for panel A,
         $n_e \!=\! 10^{24}$cm$^{-3}$ for panel B, and
         $n_e \!=\! 10^{26}$cm$^{-3}$ for panel C. 
         The thick dashed vertical line labels the ion temperature
         of $T_i \!=\! 10^5\,$K. The thin dashed line indicates
         where the degeneracy parameter $n_e\Lambda_e^3$ is unity.
         Left of this line the electrons are fully degenerate.
         \label{fig7}}
\end{figure}

Figure \ref{fig7} compares the different approximation of the FGR energy
transfer rate, i.e., Eqs.~(\ref{fgr_ad}), (\ref{fgr_ld}), and (\ref{fgr_hd})
with the full FGR expression (\ref{fgrfull}). Clearly, the reduced expression
of the FGR approach (\ref{fgr_ld}) works very well for non-degenerate systems. Here, the
expansion of the Bose function is
valid. Thus, the simple expression (\ref{fgr_ld}) can be used for all cases
with $T_e \!>\! T_i$ and nondegenerate electrons. The reduced 
expression (\ref{fgr_ad}), valid for arbitrary electron degeneracy, extents
the range of the reduced formula slightly towards smaller electron
temperatures.

However, the situation dramatically changes for cases where the electron
temperature is smaller than the ion temperature. Here, the full FGR formula 
(\ref{fgrfull}) predicts energy transfer rates that differ up to a factor of two from
the results of the reduced models. These deviations become more pronounced
with higher density and the full and reduced models show a qualitatively
different behavior for the highest density shown. The main source of deviation
is the expansion of the Bose functions $n^B_a(\omega)$ used to derive the reduced
models. As soon as the energy of the ion modes becomes comparable to the 
thermal energy of the electrons, $k_B T_e$, this expansion becomes invalid.
As the first order expansion of the Bose functions is at the heart of all
reduced descriptions, one has to solve the full FGR expression (\ref{fgr_ad})
for all cases with $T_e \!<\! T_i$ or degenerate electrons to get valid energy
transfer rates. It should be mentioned that the $T \!=\! 0$ expansion used to
derive the formula (\ref{fgr_hd}) is of course valid for low electron
temperatures. However, expression (\ref{fgr_hd}) only approaches the reduced
description (\ref{fgr_ad}) in this case and not the results from the full 
FGR description.


\section{Comparison of energy transfer from CM and FGR with MD data}

\begin{table*}[t]
 \caption{Electron-ion coupling parameter for cases A-D of Ref.~\cite{Jeon:2008}
          by Jeon {\em et al.}: 
          $n_e \!=\! n_i \!=\! 2.4 \!\times\! 10^{22}\,$cm$^{-3}$,
          $T_e \!=\! 80\,$eV, $T_i \!=\! 100\,$eV for case A;
          $n_e \!=\! n_i \!=\! 2.68 \!\times\! 10^{23}\,$cm$^{-3}$,
          $T_e \!=\! 400\,$eV, $T_i \!=\! 500\,$eV for case B;
          $n_e \!=\! n_i \!=\! 7.59 \!\times\! 10^{23}\,$cm$^{-3}$,
          $T_e \!=\! 800\,$eV, $T_i \!=\! 1000\,$eV for case C;
          $n_e \!=\! n_i \!=\! 2.4 \!\times\! 10^{25}\,$cm$^{-3}$,
          $T_e \!=\! 8000\,$eV, $T_i \!=\! 10000\,$eV for case D.
          The data for MD, MFGR$_{AA}$, MFGR$_C$, BPS, and LS$^0$ are taken from
          Ref.~\cite{Jeon:2008}. Results labelled LS$^H$ were calculated
          from Eq.~(\ref{lsh}) and TM labels data of the fit to the full
          binary collision approach obtained by Eq.~(\ref{lst}). FGR results
          were obtained from the full FGR expression (\ref{fgrfull}) and are
          identical to results from the CM approach for the conditions
          presented here. Data labelled MD$^{DD}$ are based on the Coulomb
          logarithm given by Dimonte \& Daligault \cite{Dimonte:2008}.
          The results for all cases are ordered by value.
 \label{tbl1}}
 \begin{displaymath} 
  \begin{array}{lccccccccc}
   \hline\hline
   \mbox{case} &\multicolumn{8}{c}{g_{ei}[\mbox{W}/\mbox{m}^3\!/\mbox{K}]}\\
   (\xi=R_c/\lambda_{dB})&\multicolumn{8}{c}{\mbox{ method}}\\\hline
   \mbox{A}\times10^{17}&0.88&0.90&1.03&1.31&1.42&1.51&1.52&1.53&1.57\\
   (0.58)&\mbox{MFGR}_{AA}&\mbox{LS}^0&\mbox{\bf MD}&\mbox{LS}^H&\mbox{TM}&\mbox{\bf MD}^{DD}&\mbox{FGR}&\mbox{BPS}&\mbox{MFGR}_C\\\hline
   \mbox{B}\times10^{18}&0.98&1.24&1.40&1.76&1.77&1.91&1.93&1.94&2.41\\
   (0.26)&\mbox{\bf MD}&\mbox{LS}^0&\mbox{MFGR}_{AA}&\mbox{LS}^H&\mbox{TM}&\mbox{MFGR}_C&\mbox{FGR}&\mbox{BPS}&\mbox{\bf MD}^{DD}\\\hline
   \mbox{C}\times10^{18}&3.81&4.06&4.62&5.30&5.65&5.75&5.79&5.81&7.66\\
   (0.18)&\mbox{LS}^0&\mbox{\bf MD}&\mbox{MFGR}_{AA}&\mbox{LS}^H&\mbox{TM}&\mbox{FGR}&\mbox{BPS}&\mbox{MFGR}_C&\mbox{\bf MD}^{DD}\\\hline
   \mbox{D}\times10^{20}&1.51&&1.74&1.99&2.08&2.10&2.13&2.14&3.30\\
   (0.06)&\mbox{LS}^0&-&\mbox{MFGR}_{AA}&\mbox{LS}^H&\mbox{TM}&\mbox{FGR}&\mbox{MFGR}_C&\mbox{BPS}&\mbox{\bf MD}^{DD}\\
  \hline\hline
  \end{array}
  \end{displaymath}
\end{table*}

In the literature, different quantities are applied to quantify the electron-ion
energy transfer in two-temperature systems: energy transfer rates, relaxation
times, and coupling constants. Most measures eliminate the temperature
difference, $T_e \!-\! T_i$, to allow for a more unified description although
the temperature difference might not be explicitly visible as in the LS case.

To compare with data from MD simulations, we define the electron-ion
coupling parameter $g_{ei}$
\beq
g_{ei}=g^x=\frac{Z_{ei}^x}{T_e-T_i} \,,
\label{gx}
\eeq
the arbitrary Coulomb logarithm
\beq
\ln\Lambda^x = \frac{Z_{ei}^x}{T_e-T_i} 
               \frac{m_em_i}{4\sqrt{2\pi}n_en_iZ_i^2e^4k_B}
               \left(\frac{k_BT_e}{m_e}+\frac{k_BT_i}{m_i}\right)^{3/2}
               \!\!\!\!\!\!,
\label{lnx}
\eeq
and the relaxation time
\beq
\tau^x = k_B(T_e-T_i) n_e \, \frac{1}{Z_{ei}^x} 
       = \frac{k_Bn_e}{g_{ei}} \,.
\label{taux}
\eeq
The label `$x$' denotes here the approximation level used. All three quantities
can be on the level of the full binary collision (T-matrix) approach, the
FGR expression, or the full CM description depending on the energy transfer rate
$Z_{ei}$ applied as input.

\begin{figure}[t]
\includegraphics[width=0.48\textwidth,clip=true]{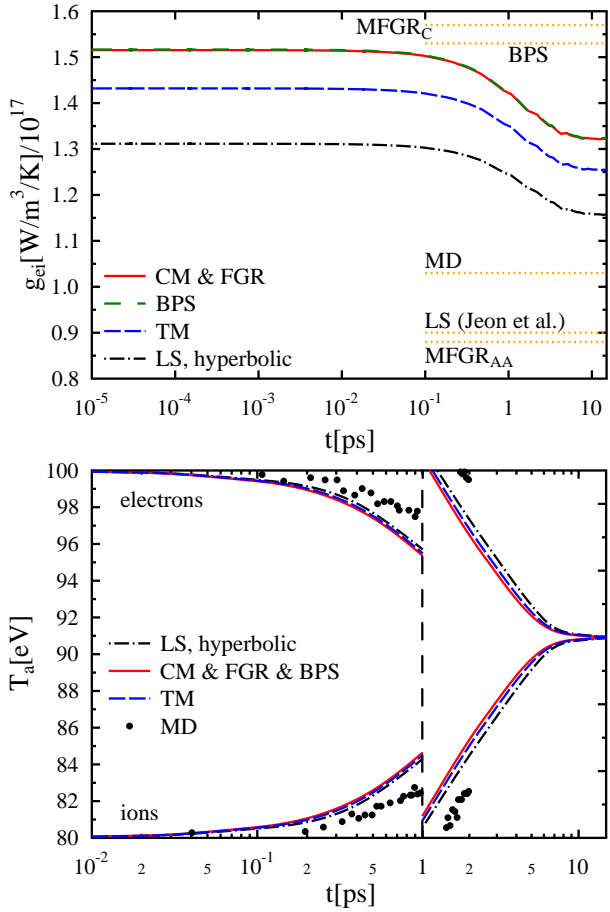}
\caption{(color online) Temperature relaxation and energy transfer rates in hydrogen
         for the case A as studied with MD in Jeon {\em et al.}
         \cite{Jeon:2008}. The plasma parameters are
         $n_e \!=\! n_i \!=\! 2.4 \!\times\!10^{22}\,$cm$^{-3}$,
         $T_e \!=\! 80$eV, and $T_i \!=\! 100$eV. The straight lines
         in the upper panel are the values as given in Ref.~\cite{Jeon:2008}.
	 {\em Beware the change in scale of the y-axis at $1$ps in the bottom panel}.
         \label{fig5}}
\end{figure}

We first compare data from MD simulations to two-temperature models employing
various energy transfer rates for cases with low degeneracy. Jeon {\em et al.}\/
provided just such data \cite{Jeon:2008}. For comparison, we add results of our
statistical models (FGR, CM) and from classical MD simulation of 
Dimonte and Daligault \cite{Dimonte:2008} in Fig.~\ref{fig5} and 
Table~\ref{tbl1}. The densities and temperatures as chosen by
Jeon {\em et al.}\/ correspond to nearly identical parameters of coupling and
degeneracy. On this basis, one would not expect qualitative changes in the
relaxation behavior between the different cases presented. Indeed, the energy
transfer rates on the level of the CM, TM, FGR, BPS, and hyperbolic LS
approaches show all a quite regular behavior: these rates keep the same order
with respect to the efficiency of the energy transfer and give consistently
a higher energy transfer than the MD simulations. The reduced FGR energy 
transfer rates MFGR$_{AA}$ as given by Jeon {\em et al.} are found however to
be smaller than MD for case A and larger than MD data for case B and C. The
modified FGR approach, MFGR$_C$ of Jeon et al., matches the full FGR formula
(blue curves in Fig. \ref{fig5}).

Interestingly, the simplest approximation to the Landau-Spitzer
formula (\ref{lnl1}), LS$^0$, using the Debye length for the upper cutoff
$b_{max}$ and the deBroglie wavelength for the lower cutoff $b_{min}$ gives
best agreement with the MD data for all cases. This might be related to
a compensation of errors. Such a result is even more surprising as 
the full FGR and TM rates incorporating Boltzmann statistics should be 
the best approximations in cases A to D. 

However, one observes that the Born parameter is smaller than unity, $\xi<1$,
and decreases further from case A to D. This fact indicates that the individual
two-particle scattering process has significant quantum diffraction effects if
it involves electrons and, thus, should be described quantum-mechanically.
Diffraction effects are approximated in the MD simulations of 
Jeon {\em et al.}\/ by using an electron-ion pseudo-potential of the
Deutsch type \cite{Jeon:2008}.

One may study the influence of the effective potential on the relaxation by
comparing with MD data obtained without the use of a pseudo-potential, e.g., by
simulating like-charges of different masses. Table~\ref{tbl1} contains such 
MD results as obtained by Dimonte \& Daligault \cite{Dimonte:2008} in addition
to the analytical results. One observes that the difference between the data
from the two different MD simulations increases from cases A to D. Moreover,
the results from MD simulations without pseudo-potential \cite{Dimonte:2008}
compare quite well to analytical theories in case A where diffraction effects 
have the least influence. 
For lower values of the Born parameter, indicating
situations with stronger diffraction effects, quantum approaches show
significantly less energy transfer in agreement with the findings of
Fig.~\ref{fig4}. This comparison indicates that the application of the 
Deutsch potential overestimates the effects of quantum diffraction. Thus,
simulations using this potential underestimate the temperature relaxation
rates for these parameters.

\begin{figure}[t]
\includegraphics[width=0.5\textwidth,clip=true]{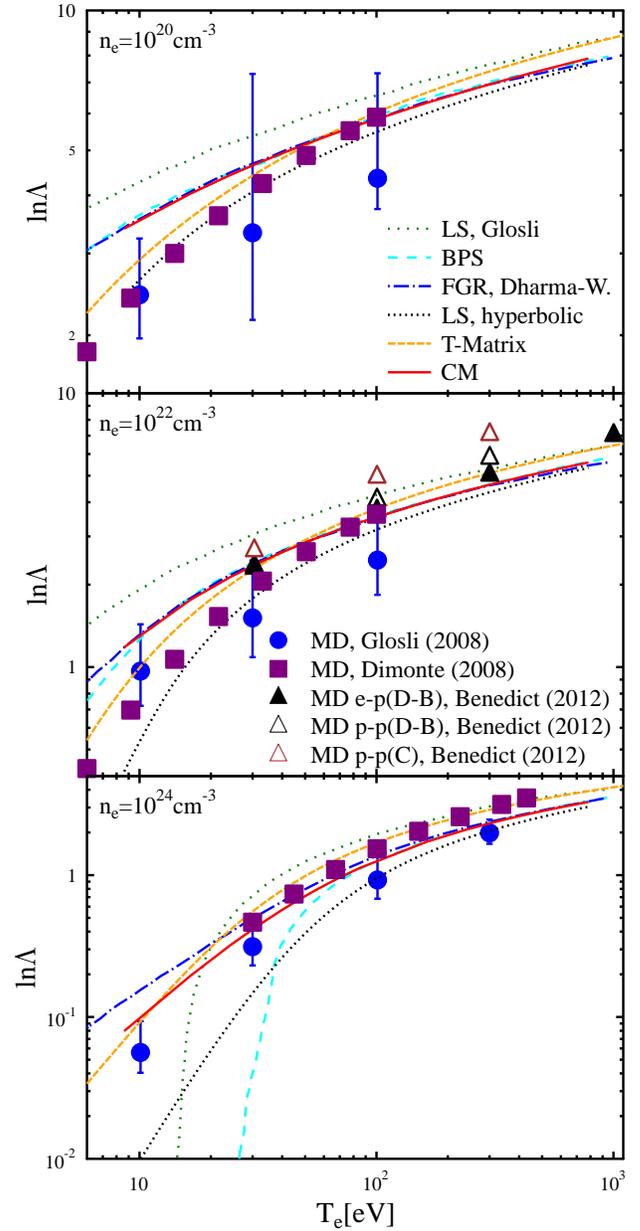}
\caption{(color online) Effective Coulomb logarithm as derived from MD simulations
         by Glosli {\em et al.}\/ \cite{Glosli:2008} and extracted
         from various other theoretical schemes for hydrogen at
         various electron densities. The ion temperature is always 
         twice as high as the electron temperature. These cases
         correspond to Tab.~I and Fig.~2 of Ref.~\cite{Glosli:2008}.
         The purple squares indicate the Coulomb logarithms from
         a fit to classical MD simulations by Dimonte \& Daligault
         \cite{Dimonte:2008}. The triangles are data from Benedict 
	  et al. \cite{Benedict:2012} \label{fig1}}
\end{figure}

\begin{figure}[!t]
\includegraphics[width=0.5\textwidth,clip=true]{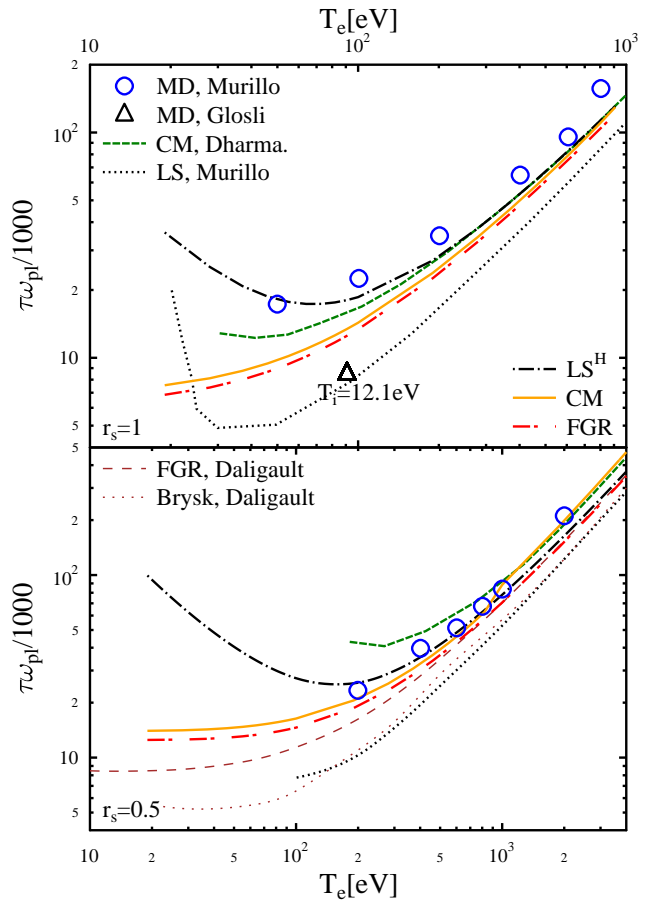}
\caption{(color online) Relaxation times obtained from MD simulations for the cases
         studied by Murillo \& Dharma-wardana \cite{Murillo:2008},
         Glosli {\em et al.}\/ \cite{Glosli:2008}, and
         Daligault \& Dimonte {\em et al.}\/ \cite{Daligault:2009}
         for a hydrogen plasmas compared to results from different
         versions of the LS, FGR, and CM approaches.
         The ion temperature is $T_i \!=\! 10\,$eV except for the
         data point from Glosli {\em et al.}.
         \label{fig6}}
\end{figure}

For high-density plasmas with lower temperatures, quantum (Fermi) statistics
and collective effects play a crucial role in the energy transfer process.
Therefore, CM and FGR energy transfer rates have been applied to adequately
describe the physics processes involved. It is demonstrated in Figs.~\ref{fig1} 
and \ref{fig6} how a description of energy transfer by MD compares against 
these theories in situations where the electrons are close to
being degenerate. 

For the two cases with lower density shown in Fig.~\ref{fig1}, there is
virtually no difference between FGR, CM, and BPS results. All these theories
predict effective Coulomb logarithms on the upper end of error bars of the
data from the MD simulations. Binary collision approaches to the electron-ion
energy transfer, namely the parametrized T-matrix model \cite{Gericke:2002b}
and the LS model featuring hyperbolic orbits, show however much better
agreement with the MD data for $n_e\!=\!10^{20}\,$cm$^{-3}$ and
$n_e\!=\!10^{22}\,$cm$^{-3}$. This fact may indicate that strong scattering as
included in the full binary collision approach \cite{Gericke:2002b} and
approximated by the LS results is important for these cases. Typically, the
Born approximation as used by the FGR and CM theories overestimates the
strength of the interaction.

The data by Glosli {\em et al.} \cite{Glosli:2008} lie also in the area
of validity for the MD results by Dimonte \& Daligault \cite{Dimonte:2008}. 
Results of the fit given by Dimonte \& Daligault and the Coulomb logarithms
extracted by Glosli {\em et al.} show good agreement for low electron
temperatures but deviations (still within the error bars given)
from $30\,$eV to $100\,$eV where quantum 
diffraction effects may become more important. The results from the fit to MD data of
Dimonte \& Daligault follow the full binary collision approach and hyperbolic
LS results more closely as found earlier. 

New extensive MD simulations by Benedict et al. \cite{Benedict:2012} 
for $n_e=10^{22}$cm$^{-3}$ show the best agreement with T-matrix results of all MD 
simulations if a Dunn\&Broyles electron-ion pseudopotential is used. When changing 
the pseudopotential and the substitution of electrons by positrons, the Coulomb 
logarithm obtained from the simulations increases and the agreement with other 
approaches worsens (see middle panel of Fig. \ref{fig1}).
The full quantum scattering theory beyond the Born approximation gives, however, 
also different cross sections if the scattering of like charges is considered. 
Thus, one may also expect different energy transfer rates.

For the highest density presented in the lower panel of Fig.~\ref{fig1}, all
variations of the LS formula as well as the BPS model break down at lower
temperatures due to the high degree of degeneracy. Differences between the
FGR approach (as calculated by Dharma-wardana \cite{Dharma:2008}) and our
CM energy transfer rate begin to show due to the occurrence of degeneracy driven ion acoustic
modes and thus CM effects \cite{Vorberger:2009}. In this case, the CM approach agrees
slightly better with the MD data. Again MD simulations of Dimonte \& Daligault avoiding
electron-ion pseudo-potentials with like-charges yield higher effective
Coulomb logarithms and, therefore, higher energy transfer rates. However, one
has to point out that classical MD simulations become questionable for
high-density systems with high degeneracy.

Fig. \ref{fig6} summarizes results for the relaxation time, as defined in
Ref.~\cite{Murillo:2008} and Eq.~(\ref{taux}), for a hydrogen plasma at even
higher densities. The results from all approaches, analytical theories and
numerical simulations, converge for high electron temperatures but differ
strongly for lower temperatures. In contrast to the findings in
Fig.~\ref{fig5}, the LS approximation using straight line trajectories and
simple cut-off parameters differs the most from the MD data whereas the
LS model using hyperbolic orbits (\ref{lsh}) matches the MD data points remarkably
well (except after the break-down at low electron temperatures). Results from
FGR models always fare worse than CM results when compared to the MD data.

The parameters investigated in Fig.~\ref{fig6} are one of the few cases where
systems with an ion temperature well below the electron temperature have been
studied by MD simulations. Under these conditions, coupled collective modes 
exist \cite{Vorberger:2009} and a deviation of the CM rates from FGR rates 
can indeed be observed. This coupled mode effect is even visible for the
highest temperatures at $r_s=0.5$. Here, the rates from the CM expression agree
with MD results and give a systematically different limit as theories not
including CM effects. Indeed, the CM theory yields approximately 30\% longer
relaxation times than FGR or LS formulas predict. 


\section{Summary and Conclusion}
We have compared data for the relaxation of two-temperature electron-ion
systems obtained by classical MD simulations with predictions of different
analytical approaches. For our extensive comparison, we considered 
i) theories treating binary collisions only (the simple LS formula and the
full binary collision (T-matrix) approach \cite{Gericke:2002b}) and 
ii) theories considering collective excitations in the plasma (FGR and
CM expressions \cite {Dharma:98}). Our main goals were to benchmark the quality
of collisional models against MD simulations and to search for indications of
coupled mode effects in the MD data. The latter should be possible as MD 
simulations not only compute the classical collision processes but 
also cover the classical dynamics and collective motions in full which in certain
parameter regimes gives rise to energy transfer via collective modes - the CM effect.

Our comparisons of MD data and theories considering binary collisions only
showed the well-known inability of the simple LS approach (considering straight
line trajectories)
to predict the electron-ion energy transfer. The LS expression for hyperbolic
orbits with appropriate cutoffs as well as the fit to the full binary collision
(T-matrix) approach \cite{Gericke:2002b} yield however satisfying agreement
with the MD data obtained by Dimonte \& Daligault \cite{Dimonte:2008} if
applied to classical plasmas. We have chosen this set of MD data here as the
underlying simulations were performed for like-charges interacting with bare 
Coulomb forces. Thus, these MD data contain no uncertainty related to the
electron-ion pseudo-potential.

In the search of CM effects, one should always compare MD data with the FGR and
CM expressions since the LS formula yields energy transfer rates and relaxation
times that may strongly depend on the cutoffs used. Therefore, deviations
between MD data and LS results are more likely related to a failure of the
LS approach than a hint of CM effects. To avoid false indications, we have also
investigated the validity of various approximations and simplifications to the
FGR expression. We found that for many cases, in particular for low electron
temperatures, none of the reduced FGR expressions are applicable as they are
all based on the expansion of the Bose function which becomes invalid. Thus,
we have evaluated the full FGR expression (\ref{fgrfull}) to make unambiguous
comparisons with the MD data.

Most MD simulations were performed for parameters where the condition for the
occurrence of ion acoustic modes in classical plasmas,
i.e.\ $Z T_e \!\gg\! T_i$, is not fulfilled. Accordingly, the data for the
electron-ion energy transfer show negligible influence of coupled collective
modes. For cases with low degeneracy where such classical simulations are
applicable, full T-Matrix and hyperbolic LS rates compare best to MD as
expected. If the electron-ion coupling is also weak, full FGR and CM rates
agree as well. For systems with degenerate electrons, neither MD simulations
nor the discussed binary collision approaches are applicable and one has,
so far, to rely on the comparison of FGR and CM rates when searching for
CM effects.

However, the MD data published in Ref.~\cite{Murillo:2008} fulfill the condition
for the occurrence of coupled electron-ion modes as the electron 
temperature exceeds the ion
temperature several times. As discussed in connection with Fig.~\ref{fig6},
the temperature relaxation times differ significantly whether or not the
theoretical approach includes coupled collective modes and the results from the
full CM description agree best with MD data. Interestingly, this relation holds
even for very high electron temperatures as predicted before
\cite{Vorberger:2009} within the weak coupling CM approach. This finding is
a strong indication for CM effects in the relaxation time data extracted from
MD simulations. A remaining source of uncertainty is the use of electron-ion
pseudo-potentials in the MD simulations. A possible overcompensation of quantum
diffraction by the effective potential will result in a reduction of the
electron-ion energy transfer and, thus, might mimic CM effects.

It has been shown that some MD simulations show strong signs of CM effects on
temperature relaxation. However, MD data must be interpreted with care as
these classical simulations do not include quantum effects. Of course, the
ions are well described by Newton's equations but the electronic 
subsystem can
exhibit both quantum degeneracy at high densities and diffraction effects 
during high-energy collisions dominating at high temperatures. The importance
of quantum effects can be estimated by the degeneracy parameter 
$n_e \Lambda^3_e$ and the Born parameter $\xi$, respectively.

\begin{figure}[t!]
\includegraphics[width=0.5\textwidth,clip=true]{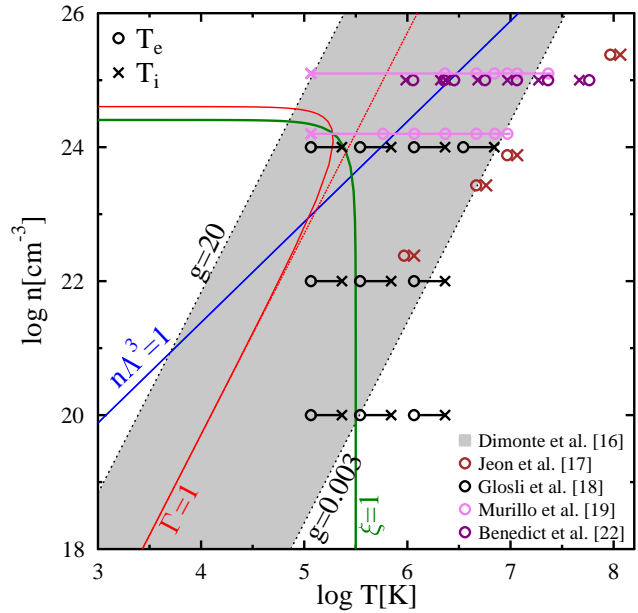}
\caption{(color online) Plasma parameters considered in recent MD simulations
         investigating electron-ion temperature relaxation in relation
         to the degeneracy parameter $n_e\Lambda_e^3$ with
         $\Lambda_e \!=\! \left(2\pi/m_e k_B T \right)^{1/2}$,
         the classical coupling parameter
         $\Gamma \!=\! \rho/d \!\sim\! \langle K\rangle/\langle V\rangle$
         with $d \!=\! (3/4\pi n)^{1/3}$),
         and the Born parameter $\xi \!=\! \rho/\lambda_{dB}$.
         Above the degeneracy line $n_e\Lambda_e^3 \!=\! 1$, the
         electrons are highly degenerate; to the right and the top of
         the curve $\xi \!=\! 1$, quantum diffractions become important
         for the two-particle scattering process. The grey area marks
         parameters considered in Ref.~\cite{Dimonte:2008}.
\label{ntplane}}
\end{figure}

In Fig.~\ref{ntplane}, the plasma parameters of MD simulations considering
two-temperature electron-ion systems have been plotted together with contours
of the degeneracy, coupling, and Born parameters. Systems above the line 
$n_e \Lambda^3_e \!=\! 1$ are plagued by quantum degeneracy while points 
on the right side of the line $\xi=1$ exhibit strong quantum diffraction that
is most likely not well described by an electron-ion pseudo-potential. To
distinguish CM effects clearly from artifacts of the pseudo-potential, we
suggest to perform MD simulations at low densities and temperatures to
maintain $n_e \Lambda^3_e \!<\! 1$ and $\xi \!<\! 1$. The need of
pseudo-potentials can be also strongly mitigated by considering systems with 
like charges. Moreover, the electrons need to be still much hotter than the
ions as CM effects occur only for $T_i \!\ll\! T_e$ in classical systems.  

Most MD simulations performed so far do not meet all requirements to clearly
observe CM mode effects as they often consider systems with $T_i \!>\! T_e$. 
Comparing the MD data with both results from a full CM theory and the FGR
approach, one finds strong indications of CM effects in the few MD data
for $T_i \!<\! T_e$. However, these simulations are too close, or even beyond,
the lines $n_e \Lambda^3_e \!<\! 1$ or $\xi \!<\! 1$ to unambiguously exclude
that the reductions in energy transfer observed are related to the application
of pseudo-potentials or the break of classical statistics inherent to
MD simulations.

\section{Acknowledgements}
The authors thank the UK's Engineering and Physical Sciences Research Council for 
financial support of this work.

\end{document}